\documentstyle[12pt]{article}
\begin{document}
%\magnification = \magstep1
\def\Ket#1{|#1 \rangle}
\def\PP{\vskip.1in}
\def\mod{{\rm mod \,} }
\centerline{THE LATTICE DYNAMICS OF COMPLETELY ENTANGLED STATES }
\centerline{ AND ITS APPLICATION TO COMMUNICATION SCHEMES}
\vskip.15in
\centerline{Daniel I. Fivel}
\vskip.15in
\centerline{ Department of Physics }
\centerline{ University of Maryland, College Park, MD 20742-4111 }
\vskip.2in
\centerline{{\bf Abstract}}
\PP
\PP
It is shown that among the orthogonal sets of EPR (completely entangled) states
there is a unique
basis (up to equivalence)  that is a also a perfectly resolved set of coherent
states with respect
to a pair of complementary  observables. This basis defines a lattice phase
space in which quadratic
Hamiltonians constructed from the observables induce site-to-site hopping at
discrete time
intervals. When recently suggested communication schemes\cite{BENa} are adapted
to the lattice they
are greatly enhanced, because the finite Heisenberg group structure allows
dynamic generation of
signal sequences using the quadratic Hamiltonians.  We anticipate the
possibility of  interferometry
by determining the relative phases between successive signals produced by the
simplest Hamiltonians
of this type, and we show that they exhibit a remarkable pattern  determined by
the number-theoretic
Legendre symbol.
\vskip 1in
\centerline{Presented at the conference on Fundamental Problems in Quantum
Theory}
\centerline{University of Maryland at Baltimore - June 1994}
 \eject
Several proposals have been made recently\cite{BENa,BENb} for employing
 EPR states\cite{EPR} in cryptographic and other communication schemes.
Evidently
 we are no longer to regard these states as theoretical curiosities useful only
for discussions of the
 foundations of quantum mechanics.  In this paper we will develop the theory of
EPR dynamics,
 showing that EPR states form a natural lattice with a finite Heisenberg group
structure, and we will
 explore the physical and mathematical consequences of that structure.

For the purposes of our discussion the term  {\it EPR states} will refer to the
subset ${\cal E}$ of
 completely entangled states in a two-particle Hilbert space  ${\cal H}$, where
each one-particle
 Hilbert space ${\cal H}_N$ is of finite dimension $N.$ It can then be
shown\cite{FIVEL} that all
 such states are of the form:
 $$
 |S\rangle = N^{-1/2}\sum\limits_{j=1}^{N}{|j,1\rangle|j^{\cal U},2\rangle},
 \eqno(1)$$
where $j$ labels any one-particle basis, and ${\cal U}$ indicates any
one particle {\it anti-unitary} transformation of the state.
     Thus,
e.g. the familiar Bohm state (the spin-0 state of two spin-1/2 particles)
is obtained in the case
 $N =2$ by choosing ${\cal U}$ to be the time-reversal transformation.
Note that the anti-unitarity of ${\cal U}$ results in $|S\rangle$ being
independent of the choice of basis so that EPR states act like unitary scalars.
Since every anti-unitary
 transformation can be obtained from one of them by applying a unitary
transformation, we may
 select a fiducial EPR state and obtain all the others from it by applying a
one-particle {\it
 unitary} operator $u$ to particle-2. Thus it makes sense to denote the set of
EPR states for given
 $N$ by $|u\rangle$ as $u$ runs over the group $U_N$ of  $N$ by $N$ unitary
matrices. If $u,v$ are
 two such matrices we  write:
 $$
 |uv\rangle = u|v\rangle,
 \eqno(2)$$
 in which $u$ on the right side acts on particle-2, and from (1) we have the
elegant formula for the
 scalar product of two EPR states:
 $$
 \langle u|v \rangle = N^{-1}Tr(u^{\dagger}v).
 \eqno(3)$$
 When we speak of orthogonal $u$'s it will be in the sense of (3). Because of
the one-one
 correspondence  we will now drop the ket notation and simply use
$u$ to denote either an EPR state
 or a one-particle operator.
{~}

There are clearly $N^2$ linearly independent EPR states, so that they span the
two particle hilbert
 space. However, {\it linear combinations of EPR states are not in general EPR
states} so that they
 themselves form a Riemannian manifold---not a subspace---in the two particle
space. Thus one produces
 new EPR states from old ones by multiplying $u$'s rather than by adding them.
This is the basis of
 the Bennett-Wiesner communication scheme\cite{BENa} which works in the
following way: Communicators
Alice   and Bob  agree on a set of $N^2$ mutually orthogonal $u$'s, say
$u_1,\cdots,u_{N^2}$ with
$u_1$ being
 the initial EPR state prepared by Alice. Any such set of $u_j$'s will be
 called an EPR-{\it basis}. Particle-2 is sent
 to Bob who applies any one of the $N^2$ operators $u_ju_1^{-1}.$  This has the
 effect of transforming {\it the two-particle state} into $u_j$, and he then
returns particle-2 to
 Alice. Since she owns both particles she can ascertain which of the $N^2$
orthogonal states the
 system is in by means of a generalized Stern-Gerlach apparatus. Now it is
important to recognize
 that Alice cannot {\it unilaterally} decide to change the initial state from
$u_1$ to one of the
 other basis states e.g.\ $u_2$ without informing Bob. The reason is that
$u_ju_1^{-1}u_2$ will not
 in general coincide (up to phase) with any one of the $N^2$ $u_j$'s. Moreover,
even if Alice does not
 change the initial state, Bob is
  still restricted to a {\it single} application of one of the operators
$u_ju_{1}^{-1}$, i.e.\ he cannot employ a dynamical process that might involve
an arbitrary product of $u_j$'s.
The reason is the same.
{~}

 Now let us observe that {\it both of the limitations just described will be
eliminated if we are able
 to choose the $N^2$ $u_j$'s in such a way that they are not only mutually
orthogonal but also
 form the ray representation of a multiplicative group} ${\cal G}$. Let us see
  that there is a choice, and (up to equivalence) a {\it unique} choice, of the
set of $u_j$'s that
 will have this property:
{~}

 We begin by observing that the qualifier ``ray" is essential in that there
will in general be no
 non-trivial {\it true} representations. To see this observe first that ${\cal
G}$ cannot be
 abelian, since with $N \times N$ matrices at most $N$ linearly independent
ones can commute. Now
 suppose that  $N = p,$ where $p$ is prime. It is known\cite{HALL} that the
{\it only} groups of
 order $p^2$ are abelian, namely  the cyclic group of order $p^2$ and the
direct product of two
 cyclic groups, each of order $p$. The first possibility does not give us a
non-abelian result even
 when we extend to a ray representation by allowing arbitrary phases. However
the  second one {\it
 does}. To see why one need only recall how the coordinate and momentum
operators $X,P$  of a
 particle each generate  abelian groups $e^{i\alpha X},e^{i\beta P}$ that also
commute with one
 another {\it except for a phase multiple}. This  immediately suggests how to
construct the solution
 in the case where $N$ is a prime and leads us to expect a finite Heisenberg
group. It
 will turn out, in fact, that the Heisenberg group is the only solution for
non-prime $N$ as well---
 provided  that the EPR states are not  composites of simpler EPR states.
 In composite cases the
 wave-function will factorize  and the associated group will be the direct
product of lower order
 heisenberg groups.
{~}

{}From the analogy with the $e^{i\alpha X},e^{i\beta P}$ operators, let
$\sigma,\tau$ be unitary
 operators in the $N$-dimensional particle-2 Hilbert space ${\cal H}_{N}^{(2)}$
 satisfying:
 $$
 \sigma \tau = \omega \tau \sigma,\qquad \omega = e^{2\pi i/N}.
 \eqno(4)$$
 These  can be realized by introducing a basis $|j\rangle \;$, $j=0,1,\cdots
(mod \; N)$ in
 ${\cal H}_{N}^{(2)}$ and letting
 $$
 \sigma|j\rangle = \omega^j |j \rangle ,\qquad \tau|j\rangle = |j+1 \rangle .
 \eqno(5)$$
 (Note that $\tau|N-1 \rangle = |0 \rangle $). We now define the $N^2$
u-matrices of our EPR basis to
 be:
   $u({\bf j})$ where ${\bf j} = (j,k)$ with $j,k =  0,1,\cdots,N-1$  and
 $$
 u({\bf j}) = e^{-i\pi jk/N}\sigma^j\tau^k.
 \eqno(6)$$
 With this choice of phases we have the simple group multiplication law:
 $$
 u({\bf j})u({\bf j'}) = e^{i\pi {\bf j \times j'}/N}u({\bf j + j'}),
  \qquad {\bf j \times j'} \equiv jk'-kj'.
 \eqno(7)$$
 This is analogous to the quantum mechanical rule for multiplying operators of
the form
 $u(\alpha,\beta) \equiv e^{i(\alpha X + \beta P)}$ and defines the {\it finite
Heisenberg group }
 ${\cal G}_N$. Note that the phase factor on the right is such that the
operators on the left commute
 whenever ${\bf j,j'}$ are linearly dependent,
 whence in particular $u(n{\bf j}) = (u({\bf j}))^n$. The trace-orthonormality
of the $u({\bf j})$'s
 may be verified by direct computation. Note that the mutual orthogonality of
the $u({\bf j})$'s
 follows from:
 $$
 Tr(u({\bf j}) ) = N\delta_{{\bf j},{\bf o}},
 \eqno(8)$$
 where ${\bf o}$ is the zero vector.
{~}

 The Heisenberg group structure we have obtained shows that the states of the
lattice are exactly
 analagous to the coherent states associated with the coordinate-momentum
Heisenberg group with the
but with discrete-valued conjugate observables $\sigma$ and $\tau$. We will
 therefore refer to the distinguished set we have constructed as a {\it
coherent EPR basis}. Note
 that the usual coherent states do not enjoy the orthogonality relationship of
the $N^2$ coherent EPR
 states. The fact that the coherent EPR basis states  are in this sense
completely {\it resolved}
 expresses the circumvention of the uncertainty principle in the
non-invasive measurements performed on EPR states.
{~}

As we have noted, the great virtue of the group property of the coherent basis
in the BW
 communication scheme is that  an arbitrary  sequence of $u$'s from the basis
can be applied, and an
 arbitrary element from the basis can be unilaterally chosen to be the initial
state. This
 flexibility also makes possible the process we describe next, namely the {\it
dynamic} generation of
 sequences  at discrete time intervals through the action of a suitable
Hamiltonian $H$. Such a
 Hamiltonian must have the property that the associated time evolution operator
$U(t)$ obtained from
 the Schr\"{o}dinger equation $i\partial U/\partial t = HU$ will transform
elements of the basis into
 one another at designated times $t = \{t_n\}.$ Thus we must require:
  $$
 U(t)u({\bf j})U^{\dagger}(t) = u({\bf j'}).
 \eqno(9)$$
 Replacing ${\bf j}$ with $n_1{\bf j_1} + n_2{\bf j_2}$ where $n_1,n_2$ are
integers and using (7),
 one deduces that ${\bf j'}$ must be related to ${\bf j}$ by a {\it linear}
transformation under
 which the cross-product is invariant. Thus each $U$ must correspond to an
element ${\cal M}$ of the
 group $SL_2(Z/N)$ consisting of  two-by-two matrices with elements in the ring
of integers $mod(N)$
 and unit determinant $(mod \; N)$.
{~}

In order to simplify the discussion without affecting the physics we will from
here on assume that
 $N$ is a prime $p$. The reason is that   $Z/p$  will then be a {\it field}
$Z_p,$ and so
 we may do  matrix  manipulations just as with complex numbers, the condition
for an
 inverse being simply that the determinant not be divisible by $p$. (No
eigenvalues will have to be
 computed, so the fact that $Z_p$ is not algebraically closed will not cause
difficulties.)
 Generalization to non-prime $N$ is straightforward.
{~}

 Our problem now is to construct  unitary operators $U_{{\cal M}}$ such that
 $$
 U_{{\cal M}}u({\bf j})U^{\dagger}_{{\cal M}} = u({\cal M}{\bf j}).
 \eqno(10)$$
 with ${\cal M} \in SL_2(Z_p)$. One guesses the answer based on  experience
with $X,P$ in
 quantum mechanics, namely that {\it quadratic} forms in $X,P$ generate linear
transformations of the
 operators. By the completeness of the $u({\bf j})$'s we know that $ U_{{\cal
M}}$
 can be expressed as a linear combination of them. We are thus led to try
linear combinations in
 which the coefficients are phases constructed by exponentiating quadratic
forms
  $\tilde{{\bf j}}{\cal Q}{\bf j}$ in which  ${\cal Q}$ is a two-by-two matrix
and
 the tilde indicates a row vector. Indeed we will find that $(10)$ can be
implemented with:
 $$
 U_{{\cal M}} = \sum\limits_{{\bf j}}{
  e^{ (\pi i/2p) \tilde{{\bf j}} {\cal Q}_{{\cal M}}
  {\bf j} } \; u({\bf j}) },
  \eqno(11)$$
 or possibly a degenerate form of this in which the double sum reduces to a
simple sum.
{~}

Let us first exploit the fact\cite{HUA} that $SL_2(Z_p)$ is generated by the
two elements ${\cal M}
 = \rho,\chi$ with:
  $$
 \rho =
 \left( \matrix{1\quad 1\hfill\cr
   0\quad 1\hfill\cr} \right), \qquad
 \chi = \left( \matrix{1\quad 0\hfill\cr
   1\quad 1\hfill\cr} \right)
 \eqno(12)$$
  One readily verifies  that the corresponding $U_{{\cal M}}$ will be the
degenerate forms of $(11)$
 alluded to. In fact,  making use of the familiar character identitiy:
   $$
 \sum\limits_{n=0}^{p-1}{e^{2\pi i n /p}} = \delta_{n0},
 \eqno(13)$$
 where the Kronecker symbol is $mod(p)$, a straightforward calculation will
verify the following
 formula for the $U_{{\cal M}}$'s corresponding to the cyclic subgroups
containing the two generators:
  $$
 {\cal M} =  \left( \matrix{1\quad m^{-1}\hfill\cr 0 \,\quad 1\hfill\cr}
\right) \;\;
  \to \;\;  U_{{\cal M}}
 =  \sum_j{e^{-  \pi i m j^2/p}u(j,0)},
 $$
 $$
 {\cal M} =
 \left( \matrix{1\qquad 0\hfill\cr  m^{-1} \;\; 1\hfill\cr} \right) \;\;
  \to \;\; U_{{\cal M}} =  \sum_k{e^{\pi i m k^2/p}u(0,k)}.
 \eqno(14)$$
 Sums are from $0$ to $p-1$. The operators
 $U_{\rho},U_{\chi}$ corresponding to the group generators  $\rho,\chi$ are
obtained by setting
 $m=1$. (Caution: In going from the left to the right side of $(14)$ the $m$
corresponding to
  $m^{-1}$ in $Z_p$ must first be  written as an {\it ordinary} integer
$mod(p)$. Thus e.g.\
 if $p = 5$ and $m^{-1} = 3$ appears in the matrix on the left, its $Z_5$
inverse $m = 1/3$ should be
 written as $m = 2$ in the exponent on the right, not directly
inserted as $m = 1/3$.)
{~}

One can now, in principle, find a $U_{{\cal M}}$ for any ${\cal M}$ by
writing it as a "word"
in the letters $ \rho $,$\chi$, replacing them with $U_{\rho }, U_{\chi }$, and
 multiplying out the $U$'s. This would be unnecessarily tedious as the
following formula indicates:
  Suppose that in
  $(11)$ we put:
 $$
 {\cal Q}_{{\cal M}} = \left( \matrix{2a \quad b \hfill\cr b \quad 2c\hfill\cr}
\right),
 $$
 with integers a,b,c and with the restrictions on the discriminant $\Delta
\equiv  b^2 - 4ac$
 $$ (i) \;\; \Delta    \not\equiv 1 (mod \; p),\qquad
  (ii) \;\; \Delta \equiv 1 \, (mod \; 4),\;\; i.e.\  b \; odd.$$
 Then it can be shown that ${\cal Q}_{{\cal M}}$ is related to ${{\cal M}}$ by
a Cayley transform:
 $$
 {\cal M} = {{\nu {\cal Q}_{{\cal M}} + I  } \over { \nu {\cal Q}_{{\cal M}} -
I }},
 \qquad \nu = \left( \matrix{\; 0 \quad \; 1\hfill\cr-1 \quad 0\hfill\cr}
\right),
 \eqno(15)$$
 where the computations in $(15)$ are in $Z_p.$ Note that condition $(i)$ on
$\Delta$
 insures that the inverse in $(13)$ exists and $(ii)$ is found to be needed in
the course
 of calculation to permit use of $(13)$.
{~}

While we shall not make use of it in the present paper it is interesting to see
from $(15)$ how the
 analogue of the canonical structure of quantum mechanics is  expressed in the
finite lattice phase
 space of EPR states: If ${\cal R}$ is a unimodular matrix one verifies that $
{\cal R}^{-1}\nu = \nu
 \tilde{{\cal R}} $ so that $(14)$ continues to  hold under the transformation:
  $$
 {\cal M} \to {\cal R}^{-1}{\cal M}{\cal R}, \qquad {\cal Q}_{{\cal M}} \to
 \tilde{{\cal R}}{\cal Q}_{{\cal M}}{\cal R}.
  \eqno(16)$$
 Thus each $U_{{\cal M}}$ producing a ``hopping" of EPR states from one lattice
site to another
 will have a counterpart under the canonical transformation ${\cal R}$ which,
as one sees from its
 relation to $\nu,$ is a finite symplectic transformation of the lattice. Note
that ${\cal R}$
 preserves the discriminant and therefore the two conditions used in $(15)$.
{~}

 Having now developed techniques for implementing $(10)$ let us next examine
the  simplest
 example of a dynamical process that will  produce a sequence of EPR states,
namely an analogue  in
 the lattice phase space of free particle motion. We produce this by iterating
the
 generator $\rho$ of the first cyclic group in $(14)$, i.e.\ we define:
  $$
 U(t) = (U_{\rho})^{2t},\quad t = 0,1,\cdots.
 \eqno(17)$$
 An effective Hamiltonian $H_{\rho}$ can be associated with this by:
 $$
 U_{\rho} = e^{-iH_{\rho}/2},
 \eqno(18)$$
 the additional factor of $2$ having been inserted anticipating a
simplification below.
 Thus $(17)$ will be the formal solution of the corresponding Schr\"{o}dinger
equation determined by
 this  Hamiltonian. A formal solution, however is not good enough for our
purposes--- we require an
 {\it explicit} formula for $U(t)$. We note that
  $$
 (U_{\rho})^{2t} = e^{i\psi (t)}U_{\rho ^{2t}},\qquad
 {\rho}^{2t} = \left( \matrix{1\quad 2t\hfill\cr 0 \quad 1\hfill\cr} \right),
 \eqno(19)$$
 and so, in view of (14), we {\it do} have an explicit formula {\it except for
the phase factor}.
   Now it is true that one does not need to know this factor for the
computation of $(9),$
 since the reciprocal appears in $U^{\dagger}$. Thus if we are only interested
in application to the
 BW communication scheme we require nothing more. However, we wish to go
further than
  merely {\it identifying} the signals, in particular we envisage the
possibility of doing some
 kind of interferometry with them. This will require a knowledge of the phases
in $(19)$, and as we
 shall see, they have a truly remarkable structure.
{~}

 Note first that from (8), taking traces on both sides of $(19)$ will give:
 $$
 e^{i\psi (t)} = p^{-1}Tr((U_{\rho}) ^{2t})
 \eqno(20)$$
 The right side has the structure of a partition function and can be
 manipulated in the same way. Thus if one inserts the right side of the first
equation in $(14)$ for
 $U_\rho$ (with $m = 1$) there will be a product of sums   indexed by
$j_1,j_2,\cdots,j_{2t}$
 containing $u(j_1 + \cdots + j_{2t},0)$ which has zero trace unless the sum of
the $j$'s is $0 \;\;
 mod(p)$. One can then use $(13)$ to pick out this term using a standard trick
and obtain:
 $$
 Tr((U_{\rho}) ^{2t}) = \sum_n{(F(n))^{2t}},
 \eqno(21)$$
 and a completion of the square gives:
 $$
  F(n) = \sum_j{e^{-i\pi  j^2/p} e^{2\pi i n j/p}} = S(-2p)e^{i\pi n^2/p}.
  \quad S(p) = \sum_n{e^{2\pi i n^2/p}}.
 \eqno(22)$$
 Thus
 $$
 e^{i\psi (t) } = p^{-1}(S(-2p))^{2t}S(p/t).
 \eqno(23)$$
 The function $S(p)$ is known as a Gauss sum and it  is fundamental in the
solution of quadratic
 diophantine equations. Let us examine the expression we have obtained a little
more closely using
 the known exact expression\cite{SCHO} for $S(p/t)$ for an arbitrary ${\it
odd}$ prime $p$ namely:
  $$
 S(p/t)=\left( {{t \over p}} \right)\sqrt {\left( {{{-1} \over p}}
\right)p}\hfill
 \eqno(24)$$
  where the Legendre symbol is defined by:
 $$
   \left( {{t \over p}} \right)=\left\{ \matrix{+1\;\hbox{if}\;t\equiv
a\;square\;\bmod \;p\hfill\cr
   -1\;\hbox{if}\;t \not\equiv a\;square\;\bmod \;p\hfill\cr} \right.
 \eqno(25)$$
 Observe that if an arbitrary constant is added to the Hamiltonian in $(18),$
$U_{\rho}$ will be
 multiplied by a phase which will in turn produce a phase to the $2t$'th power
in
 $(23)$. Hence the argument of the factor $(S(-2p))^{2t}$ can be ``gauged" away
along with the $t$
 independent phase $\sqrt{(-1/p)}$ in $(24)$. Since we know that the left side
of $(23)$ is unimodular
  it follows that the modulus of $S(-2p)$ must be just $\sqrt{p}$. Thus up to
gauge we have
 established the extremely simple fact that:
 $$
 e^{i\psi(t)} = \left( {{t \over p}} \right),
 \eqno(26)$$
 i.e.\ the time-evolving phase of the ``free" EPR state follows a pattern of
$+1$'s and $-1$'s in a
 manner with basic number theoretic significance. Just how simple this result
is may be appreciated
 as follows: The notion of ``sign" in the usual sense does not exist
 in the field  $Z_p$, but, if for  real numbers we definie a  positive
 number as one that is the square of something while a negative number is one
that isn't, then the
 notion generalizes to $Z_p$ as the Legendre symbol. It can also be shown that
that for $p>2$ there
 are just as many squares (quadratic residues) as non-squares. Thus we have
obtained the pleasing
 result that the analoge of free particle motion in the EPR lattice is
characterized by a wave
 function with the analogue of a sign alternating phase! The computation of the
Legendre symbol is
 facilitated by its well-known factorization law which reduces it to a product
of Legendre symbols
 whose upper members are the prime factors of $t$. These in turn obey the
celebrated and profound
 Gauss law of quadratic reciprocity relating $(\textstyle{q \over p})$ to
$(\textstyle{p \over q})$.
{~}
It is clear from the above discussion that when we come to investigate and
classify more general
lattice Hamiltonians we will encounter generalized gaussian sums (theta-series)
\cite{HUA,SCHO}
and  will have to invoke the general theory of quadratic diophantine equations.
It thus appears that
we have just scratched the surface of fruitful connections between the lattice
dynamics of EPR
states and one of the richest areas of contemporary mathematics. For example
 it will be of utmost interest to ascertain the quantum mechanical significance
of Gaussian quadratic
 reciprocity.
 \vskip .2in
 I should like to acknowledge useful conversations with L. Washington and A.
Dragt.
 \PP

 \end{document}